\documentclass[default]{aastex62}
\usepackage{xcolor}
\usepackage{amsmath}

\graphicspath{{./}{}}
\submitjournal{ApJ}

\definecolor{maroon}{cmyk}{0, 0.87, 0.68, 0.32}

\shorttitle{REVISITING EMPIRICAL STELLAR RELATIONS}
\shortauthors{Herbst et al.}

\begin{document}

\title{FROM STARSPOTS TO STELLAR CORONAL MASS EJECTIONS\\
-- REVISITING EMPIRICAL STELLAR RELATIONS --}

\correspondingauthor{Konstantin Herbst}
\email{herbst@physik.uni-kiel.de}

\author[0000-0001-5622-4829]{Konstantin Herbst}
\affil{Institut f\"ur Experimentelle und Angewandte Physik, Christan-Albrechts-Universit\"at zu Kiel, Leibnizstr. 11, 24118 Kiel, Germany}

\author[0000-0002-9479-8644]{Athanasios Papaioannou}
\affiliation{Institute for Astronomy, Astrophysics, Space Applications and Remote Sensing (IAASARS), National Observatory of Athens, I. Metaxa \& Vas. Pavlou St., 15236 Penteli, Greece}

\author[0000-0003-4452-0588]{Vladimir S. Airapetian}
\affiliation{NASA Goddard Space Flight Center, Greenbelt, MD, US and
American University, DC}

\author[0000-0002-9425-2123]{Dimitra Atri}
\affiliation{Center for Space Science, New York University Abu Dhabi, PO Box 129188, Abu Dhabi, UAE}

\begin{abstract}

\noindent Upcoming missions, including the James Webb Space Telescope, will soon characterize the atmospheres of terrestrial-type exoplanets in habitable zones around cool K- and M-type stars searching for atmospheric biosignatures. Recent observations suggest that the ionizing radiation and particle environment from active cool planet hosts may be detrimental for exoplanetary habitability. Since no direct information on the radiation field is available, empirical relations between signatures of stellar activity, including the sizes and magnetic fields of starspots, are often used. Here, we revisit the empirical relation between the starspot size and the effective stellar temperature and evaluate its impact on estimates of stellar flare energies, coronal mass ejections, and fluxes of the associated stellar energetic particle events. 

\end{abstract}

\keywords{Starspots -- Stellar Flares -- Coronal Mass Ejections}

\section{Introduction} \label{sec:intro}
The increasing sensitivity of ground and space-based instruments has led to an exponential growth in the detection and characterization of stellar flares. Space missions such as {\it Kepler} \citep[e.g.,][]{Koch-etal-2010}, {\it Gaia} \citep[e.g.,][]{Gaia-collab-2016} and {\it TESS} \citep[][]{Ricker-etal-2015} have contributed significantly to our understanding of the statistical properties of energies and frequencies of flares and the underlying mechanisms generating them. These new statistical studies of stellar flares complemented with multi-wavelength observations of eruptive events opened a new window to search for signatures of flare associated Coronal Mass Ejections (CMEs) and stellar energetic particle (SEP) events and their impact on exoplanetary environments \citep[e.g.,][]{Airapetian-etal-2016, Airapetian-etal-2020, Atri-2020, Herbst-etal-2019c, Scheucher-etal-2020}. Some of the most important properties of flares, CMEs, and SEPs, including their energies and frequency of occurrence, are determined by the strength of the magnetic field and the size of the stellar active region associated with the starspot area derived from observations. Thereby, the amplitude of the rotationally modulated brightness variation and the effective (photospheric) stellar temperature are directly derived from observations. In contrast, the evaluation of the starspot temperatures \citep[e.g.][]{Notsu-etal-2013} is based on the empirical scaling with the effective temperature of the star derived by \citet{Berdyugina-2005}.  Although this relation often is applied to derive the starspot areas of G-, K-, and M-dwarfs \citep[][]{Morris-etal-2018, Notsu-etal-2019, Howard-etal-2019b, Yamashiki-etal-2019}, more accurate observations and stellar targets are required for the validation and revision of the current empirical equation.

In this paper, we first reanalyze the sample by \citet{Berdyugina-2005} using the scipy.optimize.leastsq fitting routine\footnote{provided at \url{https://docs.scipy.org/doc/scipy-0.19.0/ reference/generated/scipy.optimize.leastsq.html}}. Thereby, the Jacobian matrix is multiplied with the residual variances, estimated by the mean square errors. The resulting covariance matrix is then used to derive the standard error and, therefore, the $\pm \sigma $ uncertainty. Further, we obtain a revised empirical relation, quantifying for the first time an error band.

Additionally, we expand the data set utilizing data from \citet{Biazzo-etal-2006} and \citet{Valio-2016} to derive an updated empirical relation and investigate its impact on empirical estimates of the fundamental properties of stellar flares and CMEs. 
%
\begin{figure*}[!t]
\begin{center}
 \includegraphics[width=\textwidth]{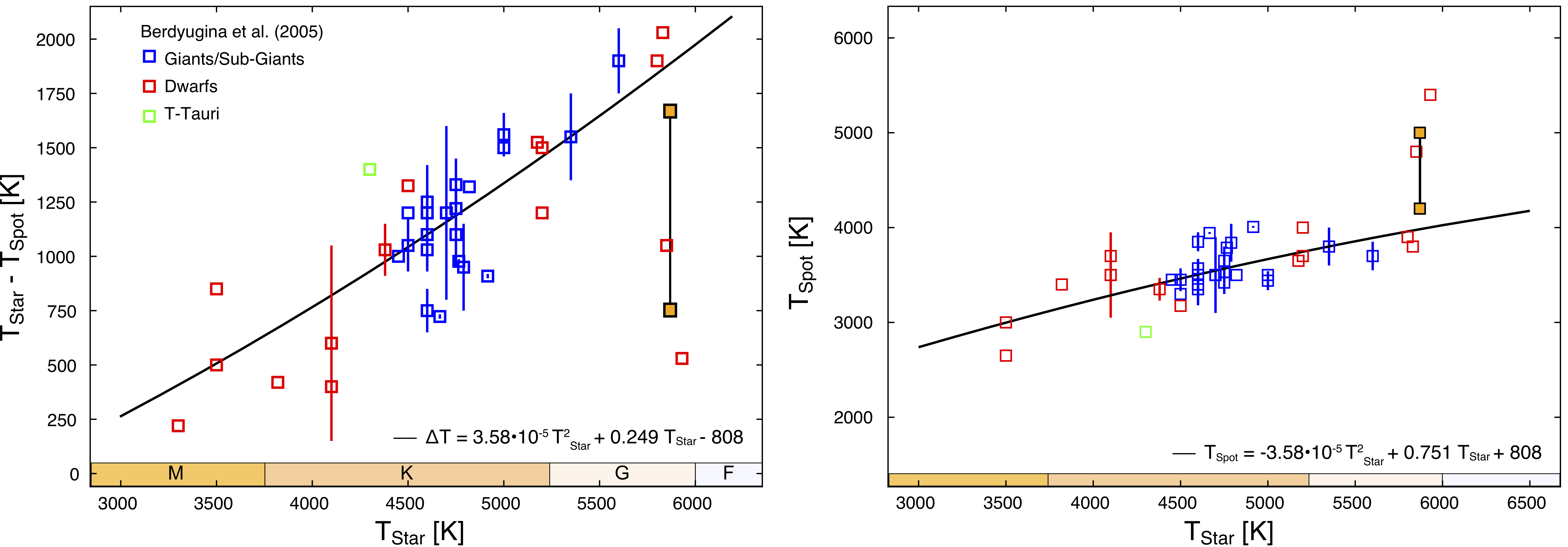}\\
  \includegraphics[width=\textwidth]{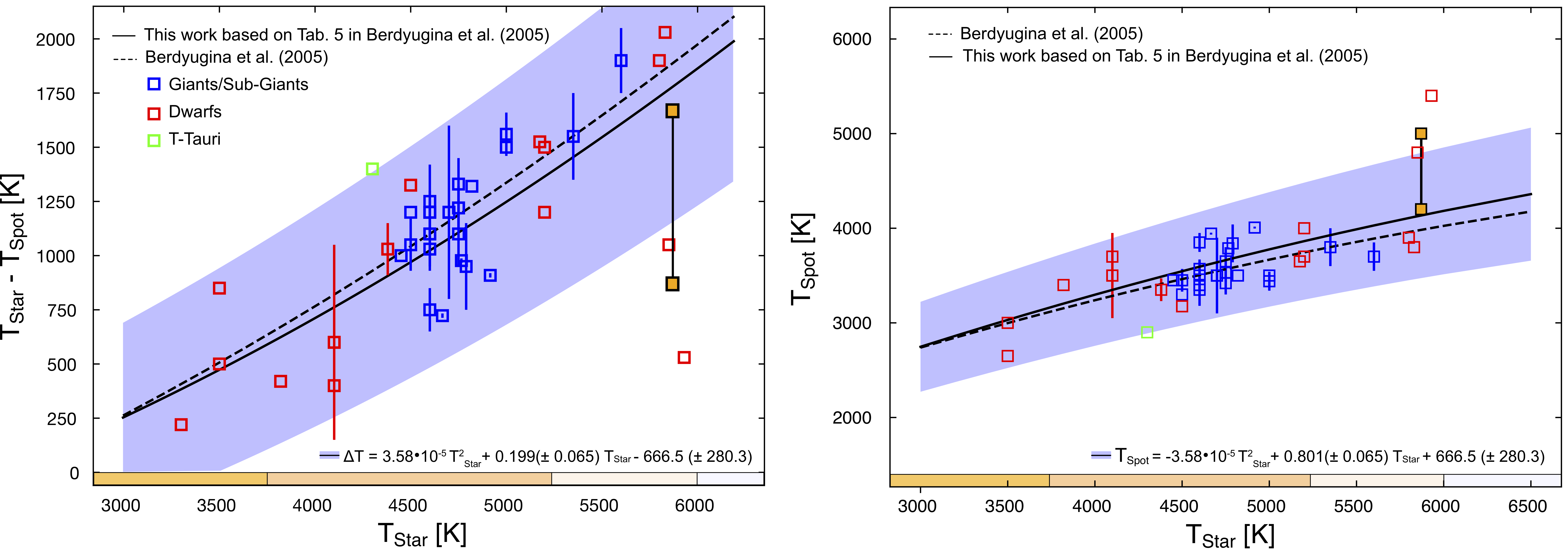}
  \caption{Upper left: Difference between stellar photosphere and spot temperatures for dwarf stars (red squares), active giants/sub-giants (blue squares), and a T-Tauri star (green square). Figure based on the data given in Table 5 of \citet{Berdyugina-2005}. Upper right: Starspot temperature as function of the stellar (photospheric) temperature. Here the the orange squares give typical solar umbra (lower symbol, $T_{\mathrm{Spot}} \approx$ 4100 K) and penumbra (upper symbol, $T_{\mathrm{Spot}} \approx$ 5000 K) temperatures. Lower left: Re-analysis of the sample discussed in \citet{Berdyugina-2005} based on the mathematical approach disscued in the text. Lower right: Corresponding $T_{\mathrm{Star}}$-$T_{\mathrm{Spot}}$-relation.}
 \label{fig:1}
 \end{center}
 \end{figure*}
%
\section{The Empirical Relation Between Stellar and Starspot Temperatures}
The empirical relation derived by \citet{Berdyugina-2005} is a second-order polynomial fit to the data representing the starspot temperature contrast with respect to the effective temperatures of cool dwarfs and giant stars. While the effective stellar temperatures are derived with high precision, the temperatures of starspots are obtained from various observational techniques including Doppler imaging, light curve modeling, and molecular bands and atomic line-depth ratios \citep[see the extended discussion in][]{Berdyugina-2005}. The spot temperature recovered from molecular band modeling uses the combining spectra of suitable standard stars of different effective temperatures weighted by a spot filling factor and continuum surface flux derived from radiative transfer models \citep[][]{Kurucz-1991} provides the continuum surface flux ratio between the spot and the quiet (unmagnetized) photosphere. Also, molecular band modeling is sensitive to the assumed photospheric temperature, chemical composition, and Doppler shifts across the spots, which may affect the spot filling factor \citep[][]{Berdyugina-2002, ONeal-etal-2004}. However, strong magnetic fields affect the optical thickness of molecular bands via modified vertical temperature gradients introduced by vertical motions (Evershed effect), which is not accounted for in "non-magnetic" radiative transfer models.

The stellar sample by \citet{Berdyugina-2005} utilizing the Doppler Imaging method includes 17 G-K (super)giants, eight main-sequence stars (two G-types and  four K- and M-types each), and one pre-main-sequence star. Interestingly, the young G-type \textit{EK Dra}, has a much lower temperature contrast than other young G-type stars and the current Sun's umbra. This may be the result of the non-uniqueness of the temperature reconstruction and the possible importance of the vertical energy transport within starspots which needs to be considered in the multi-dimensional magnetohydrodynamic modeling of magnetoconvection. 

As shown in the upper left panel of Fig.~\ref{fig:1}, \citet{Berdyugina-2005} found that the relation between the effective stellar temperature ($T_{\mathrm{Star}}$) and the starspot temperature ($T_{\mathrm{Spot}}$) for the cool star regime is best described as
\begin{eqnarray}
    \Delta T(T_{\mathrm{Star}}) =& T_{\mathrm{Star}} - T_{\mathrm{Spot}}\nonumber\\
    =& 3.58 \cdot 10^{-5}~T_{\mathrm{Star}}^2 + 0.249~T_{\mathrm{Star}}-808.\label{eq:1}
\end{eqnarray}
As can be seen, $\Delta T(T_{\mathrm{Star}})$ increases with increasing stellar temperature, varying between about 200 K for cool M-stars (M4) up to about 2000 K for G-type stars (G0), which leads to a clear tendency towards higher-contrast spots -with respect to the photosphere- in hotter stars. In addition, almost no difference can be found between active dwarfs and giants in the K- and G-star regime. The latter, according to  \citet{Berdyugina-2005}, implies that the nature of starspots most likely is the same in all active stars. 

Taking into account that $\Delta T(T_{\mathrm{Star}}) = T_{\mathrm{Star}} - T_{\mathrm{Spot}}$ the temperature of starspots can rather easily be derived once the effective stellar temperature is known. Based on Eq.~(\ref{eq:1}) the starspot temperature is given by
\begin{equation}
  T_{\mathrm{Spot}} = -3.58\cdot 10^{-5}~T_{\mathrm{Star}}^2 + 0.751~T_{\mathrm{Star}} + 808. \label{eq:2}  
\end{equation}
$T_{\mathrm{Spot}}$ as a function of the effective stellar temperature is displayed in the upper right panel of Fig.~\ref{fig:1}. 

As can be seen, starspot temperatures may vary between about 2900 K in the case of M-dwarfs up to above 4000 K on G-type stars. A direct comparison to the known sunspot (umbra)-temperature variations (filled orange squares) of about 4100 K in umbra regions and 5000 K in penumbra regions shows that the widely used relation by \citet{Berdyugina-2005} is not able to reflect the solar values. Thus, although the relation, in principle, should be applicable for all stellar types, and therefore has been widely used in the stellar and exoplanetary community \citep[see, e.g.,][]{Notsu-etal-2013, Maehara-etal-2017}, it should be treated with caution, in particular in studies of young G-type stars \citep[most recently, e.g.,][]{Notsu-etal-2019, Namekata-etal-2019, Howard-etal-2019b}.
%
\subsection{Re-analysis of the sample by \citet{Berdyugina-2005}}\label{sec:approach}
Figure 5 of \citet{Berdyugina-2005} presents no error estimates for the relation between the effective stellar temperature and the temperature of the starspot, thus, for the first time, our approach allows us to estimate an error band. We find the relation between the effective stellar and the starspot temperature to be best described by:
\begin{equation}
\begin{split}
\Delta T(T_{\mathrm{Star}}) =~&3.58 \cdot 10^{-5}~T_{\mathrm{Star}}^2 + 0.199(\pm 0.065)~T_{\mathrm{Star}}\\
&-666.5 (\pm 280.3),\label{eq:3}
\end{split}
\end{equation}
as shown in the lower left panel of Fig.~\ref{fig:1}. In addition, the corresponding relation between stellar temperature and starspot temperature then is given by:

\begin{equation}
\begin{split}
T_{\mathrm{Spot}} =~& - 3.58 \cdot 10^{-5}~T_{\mathrm{Star}}^2 + 0.801(\pm 0.065)~T_{\mathrm{Star}}\\
&+ 666.5 (\pm 280.3)\label{eq:4}
\end{split}
\end{equation}
as shown in the lower right panel of Fig.~\ref{fig:1}. Both panels show a direct comparison between the stellar sample (colored symbols), the relations by \citet{Berdyugina-2005} (Eqs.~(\ref{eq:1}) and (\ref{eq:2}), dashed lines), and the newly-derived relations given by Eqs.~(\ref{eq:3}) and (\ref{eq:4}), displayed as solid lines with corresponding blue error band.

As can be seen, Eq.~(\ref{eq:1}) is well within the error band of our newly-derived relation. However, although being in good agreement in the cool M-star regime Eqs.~(\ref{eq:1}) and (\ref{eq:3}) differ by up to 169$\pm$683 K in the case of G-type stars. Nevertheless, the newly-derived relationships, including their $\pm \sigma$ uncertainties, are in good agreement with the stellar sample discussed in \citet{Berdyugina-2005}. In particular, Eq.~(\ref{eq:4}) suggests that in the case of Sun-like stars (with $T_{\mathrm{Star}}$ = 5780 K) starspot temperatures of $4100\pm 656$ K can be expected. This is also in good agreement with the range of the measured solar umbra/penumbra temperatures. 
%
\subsection{Extending the stellar sample by \citet{Berdyugina-2005}}\label{sec:newsample}
\begin{deluxetable}{c c c c}[!t]
    \tabletypesize{\footnotesize}
    \tablecolumns{4}
    \tablecaption{Comparison of former (second column) and newly-derived values of $T_{\mathrm{Spot}}$ based on Eq.~\ref{eq:4} (third column) and Eq.~\ref{eq:5} (fourth column). The former derived values are based on \citet{Berdyugina-2005} (see Eq.~(\ref{eq:2})).}\label{tab:1}
    \tablehead{\colhead{$T_{\mathrm{eff}}$ [K]} & \colhead{$T_{\mathrm{Spot}}$ [K]} & \colhead{$T_{\mathrm{Spot}} \pm \Delta T_{\mathrm{Spot}}$ [K]} & \colhead{$T_{\mathrm{Spot}} \pm \Delta T_{\mathrm{Spot}}$ [K]}}
    \startdata
       3000 & 2739 & 2747 $\pm$  475 & 2495 $\pm$ 498\\
       3200 & 2845 & 2863 $\pm$  488 & 2654 $\pm$ 510\\
       3400 & 2948 & 2976 $\pm$  501 & 2811 $\pm$ 523\\
       3600 & 3048 & 3086 $\pm$  514 & 2964 $\pm$ 535\\
       3800 & 3145 & 3193 $\pm$  527 & 3115 $\pm$ 547\\
       4000 & 3239 & 3297 $\pm$  540 & 3263 $\pm$ 559\\
       4200 & 3331 & 3399 $\pm$  553 & 3408 $\pm$ 571\\
       4400 & 3419 & 3498 $\pm$  566 & 3550 $\pm$ 583\\
       4600 & 3505 & 3594 $\pm$  579 & 3690 $\pm$ 595\\
       4800 & 3588 & 3686 $\pm$  592 & 3826 $\pm$ 607\\
       5000 & 3668 & 3777 $\pm$  605 & 3960 $\pm$ 619\\
       5200 & 3745 & 3864 $\pm$  618 & 4090 $\pm$ 631\\
       5400 & 3819 & 3948 $\pm$  631 & 4218 $\pm$ 643\\
       5600 & 3891 & 4029 $\pm$  644 & 4343 $\pm$ 655\\
       5800 & 3959 & 4108 $\pm$  657 & 4465 $\pm$ 667\\
       6000 & 4025 & 4184 $\pm$  670 & 4585 $\pm$ 679\\
       6200 & 4088 & 4257 $\pm$  683 & 4701 $\pm$ 691\\
    \enddata
\end{deluxetable}
By now two more samples have been published in the literature: 
\begin{enumerate}
\item \citet{Biazzo-etal-2006}, who investigated the observations of three K-type stars IM Peg (K2), HK Lac (K0), and VY Ari (K3-4) and the three G-type stars $\lambda$ And (G8), $\kappa$1 Cet (G5), and HD 206860
\item \citet{Valio-2016}, who modeled the G-type stars CoRoT-2 (G7), CoRoT-18 (G9), Kepler-17 (G2), and Kepler-63 (G8), the F-type stars CoRoT-4 (F8), CoRoT-5 (F9), and CoRoT-6 (F9) as well as the K-star CoRoT-8 (K1) based on a planetary transit model.
\end{enumerate}
Both samples have been derived by methods other than the Doppler Imaging: while \citet{Biazzo-etal-2006} based their analysis on photospheric line-depth ratios and utilized the simplified spot model by \citet{Frasca-etal-2005}, \citet{Valio-2016} used an exoplanet transit model to investigate the stellar relation between $T_{\rm{Spot}}$ and $T_{\rm{Star}}$. The spatial resolution of both the Doppler Imaging and the transit method is poor. In contrast, no information on both the spatial resolution and the filling factors the starspots have in active regions is available utilizing the method discussed in \citet{Biazzo-etal-2006}. Consequently, in all three cases, the resulting spot contrasts should only be seen as typical averaged values that might be underestimated. Thus, further observations, for example, using instruments like the \textit{XMM-Newton} and \textit{NICER}, as well as numerical modeling of starspot temperatures of cool dwarfs allowing for an extended empirical characterization, are required.

\begin{figure}[!t]
\begin{center}
 \includegraphics[width=0.6\textwidth]{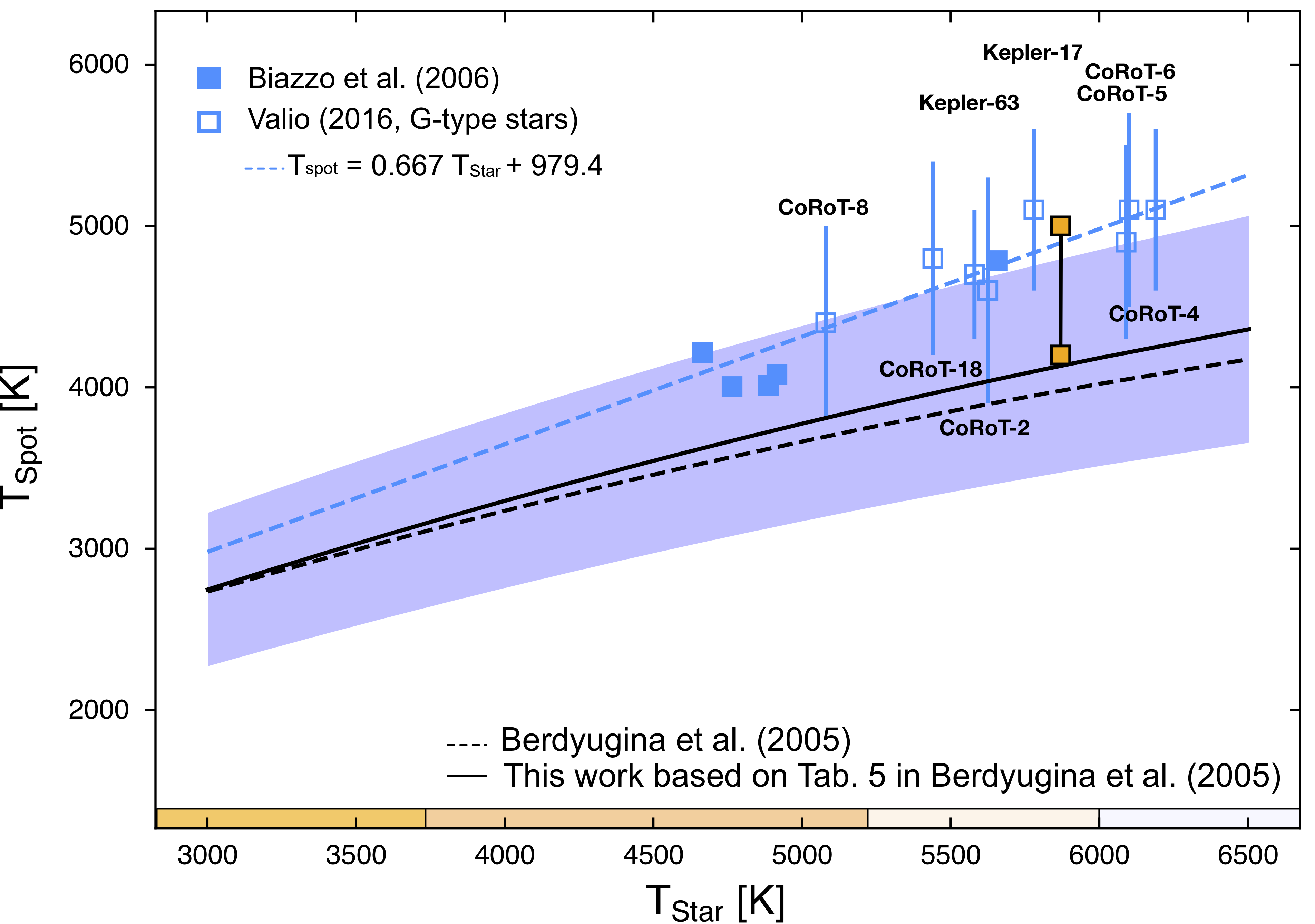}\\
 \includegraphics[width=0.6\textwidth]{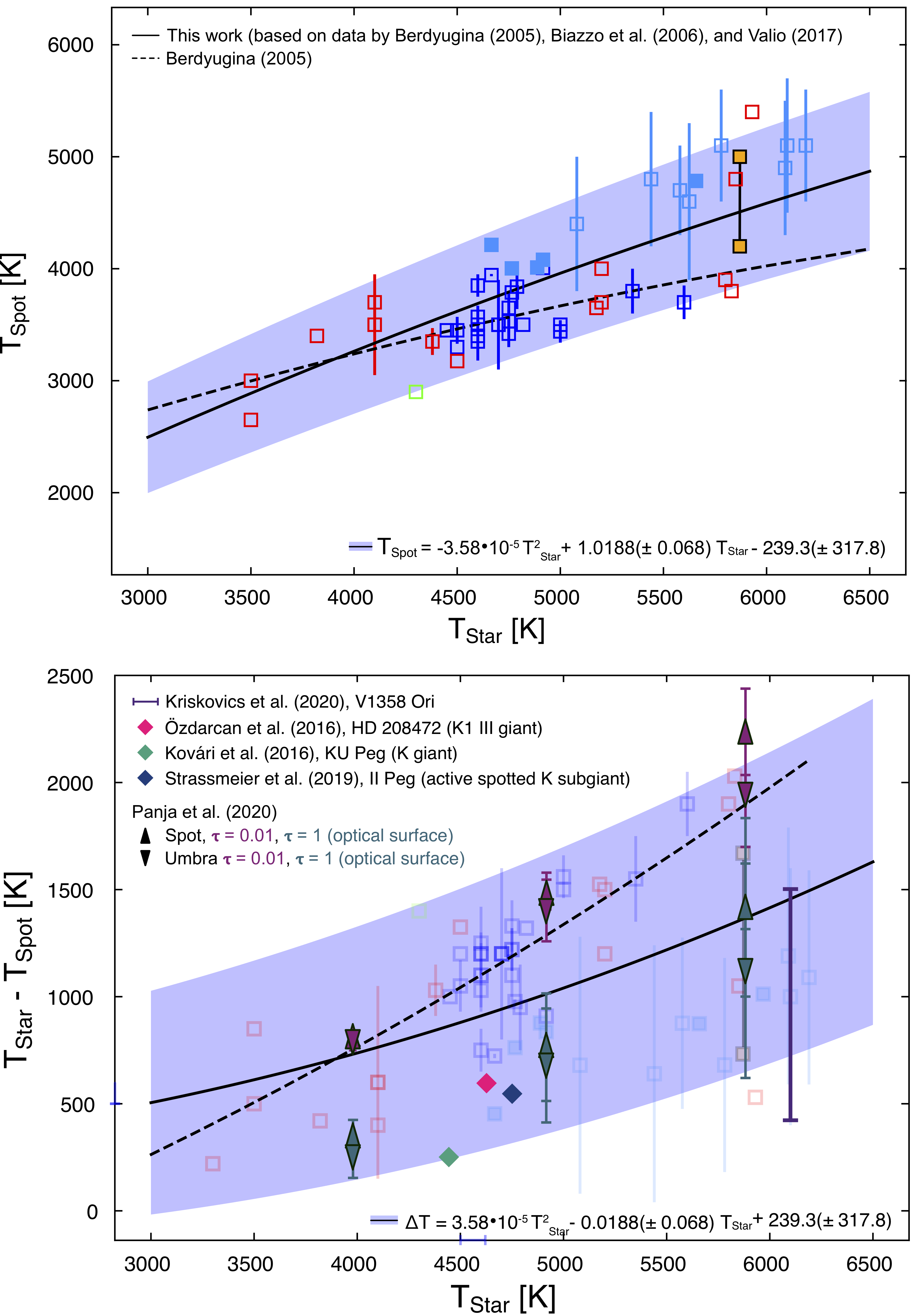}
  \caption{Upper panel: Comparison of Eq.~(\ref{eq:4}) to the sample by \citet{Biazzo-etal-2006} and \citet{Valio-2016}. Middle panel: Newly-derived $T_{\mathrm{Star}}$-$T_{\mathrm{Spot}}$-relation based on the entire sample. Lower panel: Corresponding newly-derived $T_{\mathrm{Star}}$-$\Delta T(T_{\mathrm{Star}})$-relation.}
 \label{fig:2}
 \end{center}
 \end{figure}

A comparison of both samples to the relation between stellar and starspot temperatures based on Eqs.~(\ref{eq:2}) and (\ref{eq:4}) is given in the upper panel of Fig.~\ref{fig:2}. It shows that the relation by \citet{Berdyugina-2005} given as a dashed line is not able to reflect the starspot temperatures of both \citet{Biazzo-etal-2006} and \citet{Valio-2016}, which strengthens the argument that the function needs to be treated with caution when G-type stars are investigated.  However, our newly-derived function (Eq.~(\ref{eq:4}), solid black line), and its error band is in relatively good agreement with the new sample. Please note that in addition the linear fit provided by \citet{Valio-2016} is provided (blue dashed line) which is best represented as $T_{\mathrm{Spot}} = 0.667 \cdot T_{\mathrm{Star}} + 979.4$, which well describes the new sample, however, can not characterize the stellar sample by \citet{Berdyugina-2005}.

Assuming the nature of starspots to be the same in all active stars a more reliable correlation that is able to describe the data of all three samples is needed. Therefore, in a second step, we:
\begin{enumerate}
    \item[1.] extended the sample used by \citet{Berdyugina-2005} by including the data from \citet{Biazzo-etal-2006} and \citet{Valio-2016},
    \item[2.] utilized the same approach discussed in Sec.~\ref{sec:approach} to derive a more universal relations for both $T_{\mathrm{Spot}}$ and $\Delta T(T_{\mathrm{Star}})$.
\end{enumerate}
Based on the samples by \citet{Berdyugina-2005}, \citet{Biazzo-etal-2006}, and \citet{Valio-2016} we find $T_{\mathrm{Spot}}$ to be represented best as
\begin{equation}
\begin{split}
T_{\mathrm{Spot}} = ~&- 3.58 \cdot 10^{-5}~T_{\mathrm{Star}}^2 + 1.0188(\pm 0.068)~T_{\mathrm{Star}}\\
&+ 239.3 (\pm 317.8).\label{eq:5}
\end{split}
\end{equation}
A comparison between the sample (colored symbols), the originally derived function by \citet{Berdyugina-2005} (dashed line) and Eq.~(\ref{eq:5}) is shown in the middle panel of Fig.~\ref{fig:2} and the third column of Table~\ref{tab:1}. As can be seen, Eq.~(\ref{eq:5}) represents the entire stellar sample, including, for the first time, the well-known solar sunspot temperature variations.

The lower panel of Fig.~\ref{fig:2} shows the results for the difference between stellar and starspot temperature which according to Eqs.(\ref{eq:1}) and (\ref{eq:5}) is given by:
\begin{equation}
\begin{split}
\Delta T(T_{\mathrm{Star}}) =~& 3.58 \cdot 10^{-5}~T_{\mathrm{Star}}^2 - 0.0188(\pm 0.068)~T_{\mathrm{Star}}\\ &- 239.3 (\pm 317.8).\label{eq:6}
\end{split}
\end{equation}
As can be seen, $\Delta T(T_{\mathrm{Star}})$ still increases with increasing stellar temperature, as indicated by \citet{Berdyugina-2005}, however, is less steep than previously suggested. In the mean $\Delta T(T_{\mathrm{Star}})$ varies between about 500 K for cool M-stars (M4) up to about 1500 K for G-type stars (G0). Consequently, the higher-contrast of spots with respect to the photosphere of hotter stars is not necessarily as pronounced as previously thought.
%
%
In order to test the validity of Eq. (\ref{eq:6}) a comparison with the most recent measurements and simulations is performed. Shown are the results of the Doppler Imaging of V1358 Ori, a young G9-type star \citep[purple line][]{Kriskovics-etal-2019}, the K-giants HD208472 \citep[magenta diamond ][]{Oezdarcan-etal-2016} and KU Peg \citep[green diamond][]{Kovari-etal-2016}, the active K subgiant II Peg \citep[blue diamond][]{Strassmeier-etal-2019}, and the new 3D MHD model results by \citet{Panja-etal-2020} (colored triangles). As can be seen, both the observations and the simulations are remarkably well-described by our newly-derived relation. In addition, the results by \citet{Panja-etal-2020} suggest a systematically lower contrast of about 300-400 K compared to that derived by \citet{Berdyugina-2005}.

More importantly, $\Delta T(T_{\mathrm{Star}})$ is often used as a semi-empirical quantity to derive other stellar relations like, for example, to estimate starspot areas \citep[see, e.g.,][]{Notsu-etal-2013, Shibata-etal-2013, Notsu-etal-2019}, stellar flare energies \citep[see, e.g.,][]{Notsu-etal-2019, Howard-etal-2019b}, and the properties of stellar Coronal Mass Ejections \citep[CMEs, see, e.g.,][]{Takahashi-etal-2016}. Thus, the newly-derived functions discussed here have a direct impact on previously published results and the semi-empirical relations based upon these functions.

%
\begin{deluxetable*}{c c c c c c c }[!t]
    \tabletypesize{\footnotesize}
    \tablecolumns{10}
    \tablecaption{Derived starspot temperatures $T_{\mathrm{Spot}}$, starspot areas normalized to the area of the star $A_{\mathrm{Spot}}/A_{\mathrm{Star}}$ for selected stellar \textit{Kepler} observations of M-, K, and G-type stars published in  \citet{McQuillan-etal-2014}. The calculated values based on the relations published by \citet{Berdyugina-2005} are given in columns 4 and 6, respectively. The results based on the newly-derived relations (Eqs. (\ref{eq:5}), and (\ref{eq:6})) are given in columns 5 and 7. (This table is available in its entirety in a machine-readable form in the online journal. A portion is shown here for guidance regarding its form and content.)
    \label{tab:2}}
    \tablehead{\colhead{KIC ID} & \colhead{$T_{\mathrm{Star}}$\tablenotemark{a} [K]} & \colhead{$\Delta F/F$\tablenotemark{a}}& \colhead{$T_{\mathrm{Spot}}$\tablenotemark{b} [K]}& \colhead{$T_{\mathrm{Spot}} \pm \Delta $\tablenotemark{c} [K]}& \colhead{$A_{\mathrm{Spot}/A_{\mathrm{Star}}}$\tablenotemark{b}} & \colhead{$A_{\mathrm{Spot}/A_{\mathrm{Star}}}\pm \Delta $\tablenotemark{c}}
    }
    \startdata
\hline
6060042 & 3300 & 0.011 & 2896 & 2733 $\pm$ 542 & 0.027 & 0.020$\substack{+0.340\\-0.007}$\\
10067340 & 3400 &  0.023 & 2948 & 2811 $\pm$ 549 & 0.052 & 0.042$\substack{+0.446\\-0.014}$\\
5193112 & 3500 &  0.009 & 2998 & 2888 $\pm$ 556 & 0.019 & 0.016$\substack{+0.122\\-0.005}$\\
8183277 & 3600 &  0.006 & 3048 & 2964 $\pm$  563 & 0.012 & 0.010$\substack{+0.061\\-0.003}$\\
5428199 & 3700 &  0.002 & 3097 & 3040 $\pm$ 570 & 0.004 & 0.004$\substack{+0.020\\-0.001}$ \\
2832398 & 3800 &  0.017 & 3145 & 3115 $\pm$ 576 & 0.032 & 0.031$\substack{+0.124\\-0.010}$ \\
6584163 & 3900 &  0.006 & 3192 & 3190 $\pm$ 583 & 0.010 & 0.011$\substack{+0.036\\-0.003}$ \\
3974567 & 4000 &  0.012 & 3239 & 3263 $\pm$ 590 & 0.020 & 0.021$\substack{+0.062\\-0.006}$ \\
5513890 & 4100 &  0.005 & 3285 & 3336 $\pm$ 597 & 0.008 & 0.008$\substack{+0.022\\-0.002}$ \\
3954449 & 4200 &  0.012 & 3331 & 3408 $\pm$ 604 & 0.021 & 0.022$\substack{+0.053\\-0.007}$ \\
4764328 & 4300 &  0.006 & 3375 & 3480 $\pm$ 610 & 0.009 & 0.010$\substack{+0.021\\-0.003}$ \\
6225454 & 4400 &  0.011 & 3419 & 3550 $\pm$ 617 & 0.018 & 0.020$\substack{+0.038\\-0.006}$ \\
7867109 & 4500 &  0.006 & 3463 & 3620 $\pm$ 624 & 0.009 & 0.011$\substack{+0.019\\-0.003}$ \\
3337674 & 4600 &  0.050 & 3505 & 3690 $\pm$ 630 & 0.075 & 0.085$\substack{+0.140\\-0.023}$ \\
4842623 & 4700 &  0.012 & 3547 & 3758 $\pm$ 638 & 0.018 & 0.021$\substack{+0.032\\-0.006}$ \\
2711354 & 4800 &  0.006 & 3588 & 3826 $\pm$ 644 & 0.009 & 0.011$\substack{+0.015\\-0.003}$ \\
3763109 & 4900 &  0.004 & 3628 & 3893 $\pm$ 651 & 0.006 & 0.007$\substack{+0.009\\-0.002}$ \\
3834448 & 5000 &  0.003 & 3668 & 3960 $\pm$ 658 & 0.004 & 0.004$\substack{+0.005\\-0.001}$ \\
3097429 & 5100 &  0.015 & 3707 & 4025 $\pm$ 665 & 0.021 & 0.025$\substack{+0.029\\-0.006}$ \\
2993430 & 5200 &  0.005 & 3745 & 4090 $\pm$ 672 & 0.007 & 0.008$\substack{+0.009\\-0.002} $\\
3628249 & 5300 &  0.002 & 3783 & 4155 $\pm$ 678 & 0.003 & 0.004$\substack{+0.004\\-0.001}$ \\
4077780 & 5400 &  0.016 & 3819 & 4218 $\pm$ 685 & 0.022 & 0.026$\substack{+0.025\\-0.006}$ \\
4276419 & 5500 &  0.015 & 3856 & 4281 $\pm$ 692 & 0.020 & 0.024$\substack{+0.021\\-0.005}$ \\
5531826 & 5600 &  0.010 & 3891 & 4343 $\pm$ 699 & 0.013 & 0.016$\substack{+0.014\\-0.004}$ \\
5898935 & 5700 &  0.021 & 3926 & 4405 $\pm$ 705 & 0.027 & 0.033$\substack{+0.027\\-0.007}$ \\
4069078 & 5800 &  0.009 & 3959 & 4465 $\pm$ 713 & 0.011 & 0.013$\substack{+0.010\\-0.003} $\\
3343950 & 5900 &  0.005 & 3993 & 4525 $\pm$ 719 & 0.007 & 0.008$\substack{+0.006\\-0.002}$ \\
3547642 & 6000 &  0.002 & 4025 & 4585 $\pm$ 726 & 0.002 & 0.002$\substack{+0.002\\-0.001}$ \\
5282952 & 6100 &  0.001 & 4057 & 4643 $\pm$ 733 & 0.002 & 0.002$\substack{+0.001\\-0.001}$ \\
1570150 & 6200 &  0.011 & 4088 & 4701 $\pm$ 740 & 0.014 & 0.017$\substack{+0.011\\-0.003}$ \\
5196646 & 6300 &  0.001 & 4118 & 4758 $\pm$ 746 & 0.001 & 0.001$\substack{+0.001\\-0.001}$ \\
7116137 & 6400 &  0.002 & 4148 & 4815 $\pm$ 753 & 0.003 & 0.003$\substack{+0.002\\-0.001}$ \\
\enddata
\tablenotetext{a}{data by \citet{McQuillan-etal-2014}}
\tablenotetext{b}{based on the relations according to \citet{Berdyugina-2005} (Eqs.~(\ref{eq:1}) and (\ref{eq:2}))}
\tablenotetext{c}{this work, based on the newly-derived relations given in Eqs.~(\ref{eq:5}) and (\ref{eq:6})}
\end{deluxetable*}
\section{From Starspot Temperatures to the Estimation of the Starspot Size}
According to \citet{Maehara-etal-2012}, \citet{Shibata-etal-2013}, and \citet{Notsu-etal-2013, Notsu-etal-2019} the stellar area covered by a starspot ($A_{\mathrm{Spot}}$) normalized by the projected stellar hemispherical area $A_{\mathrm{Star}}= \pi \cdot R_{\mathrm{Star}}^2$ can be described as
\begin{equation}
    \frac{A_{\mathrm{Spot}}}{A_{\mathrm{Star}}} = \left[1-\left(\frac{T_{\mathrm{Spot}}}{T_{\mathrm{Star}}}\right)^4\right]^{-1} \frac{\Delta F}{F},\label{eq:aspot_2013}
\end{equation}
where  $\Delta F/F$ is the stellar brightness variation normalized by the average stellar brightness measured by, for example, the \textit{Kepler} instrument \citep[see, e.g.,][]{Shibayama-etal-2013, McQuillan-etal-2014}, the \textit{TESS} instrument \citep[see, e.g.,][]{Stassun-etal-2019}, or the \textit{Evryscope} instrument \citep[see][]{Howard-etal-2019a, Howard-etal-2019b}. 

To investigate the impact of the newly-derived starspot-temperature-relation given in Eq.~(\ref{eq:5}), Table~\ref{tab:2}, amongst others, shows the derived $A_{\rm{Spot}}/A_{\rm{Star}}$ values of selected \textit{Kepler} main-sequence stars published in \citet{McQuillan-etal-2014} with stellar temperatures between 3300 K (mid M-stars) and 6400 K (late G-stars). Additionally, the stellar brightness variation normalized by the average stellar brightness ($\Delta F/F$) and the starspot temperatures derived based on Eq. (\ref{eq:2}) (fourth column) and Eq. (\ref{eq:5}) (fifth column) are listed.  

It shows that the derived starspot temperatures differ on average by 160 K in the case of M-stars and up to 660 K in G-type stars. More importantly, utilizing the whole stellar sample by \citet{McQuillan-etal-2014}, the average difference of the derived $A_{\rm{Spot}}/A_{\rm{Star}}$-values based on the former relations by \citet{Berdyugina-2005} and the newly-derived relation discussed in this study is in the order of 20\% in the case of G-type stars while being up to 40\% in the case of M-type stars, implying that previous studies most likely underestimated the sizes of, in particular, starspots on M-stars.

Assuming the existence of a few large starspot groups on the stellar surface and starspot lifetimes exceeding the rotational period of the star \citep[][]{Maehara-etal-2017}, based on Eqs.~(\ref{eq:1}), (\ref{eq:5}) and (\ref{eq:aspot_2013}) the area of the starspots normalized to the area of the solar hemisphere ($A_{\odot}$) further can be expressed as
\begin{equation}
 A_{\mathrm{Spot}} = \left(\frac{R_{\mathrm{Star}}}{R_{\odot}}\right)^2 \frac{T_{\mathrm{Star}}^4}{T_{\mathrm{Star}}^4 - \left[T_{\mathrm{Star}} - \Delta T (T_{\mathrm{Star}})\right]^4} \frac{\Delta F}{F_{\mathrm{Star}}}.\label{eq:aspot}
\end{equation}
A comparison of the solar spot group areas \citep[black dots, ][]{Sammis-etal-2000} and the 30-min cadence sample of Sun-like stars  ($T_{\rm{Star}}$=5100 - 6000 K) by \citet{Notsu-etal-2019}(blue crosses) is displayed in the upper panel of Fig.~\ref{fig:3}. As already discussed in \citet{Notsu-etal-2019}, the majority of the detected superflares seem to occur on Sun-like stars with large starspots \citep[see discussion in][]{Notsu-etal-2019}. In comparison, the values based on the newly derived relations are shown in the middle panel of Fig.~\ref{fig:3}. Note that the original sample by \citet{Notsu-etal-2019} has further been updated by utilizing the stellar temperatures and stellar radii given by the Gaia-DR2 catalog \citep{Berger-etal-2018}. Although the derived stellar spot group areas are in fairly good agreement, our approach, for the first time, gives an error estimate depending on the characteristics of each star. As in the case of the $A_{\rm{Spot}}/A_{\rm{Star}}$ values derived from the sample by \citet{McQuillan-etal-2014}, our newly-derived $A_{\rm{Spot}}/A_{\rm{Sun}}$ values based on the updated sample by \citet{Notsu-etal-2019} indicate a previous underestimation of the starspot sizes of Sun-like stars.

As discussed, the latter, however, has a more substantial impact on the K- and M-star regime. Based on the \textit{Evryscope} light curve measurements of 113 cool stars with intense stellar flares \citep{Howard-etal-2019a, Howard-etal-2019b} new spot group areas have been derived. The results are shown in the lower panel of Fig.~\ref{fig:3}. As can be seen, the newly-derived upper limits result in group areas which are up to one order of magnitude higher than the previously derived values \citep[see][Table 1]{Howard-etal-2019b}, which has further implications on the interpretation of the corresponding flare energies possibly released from these starspots.
%
\section{FROM Starspot Sizes To Stellar Flare energies}
\begin{figure}[!t]
\begin{center}
    \includegraphics[width=0.48\textwidth]{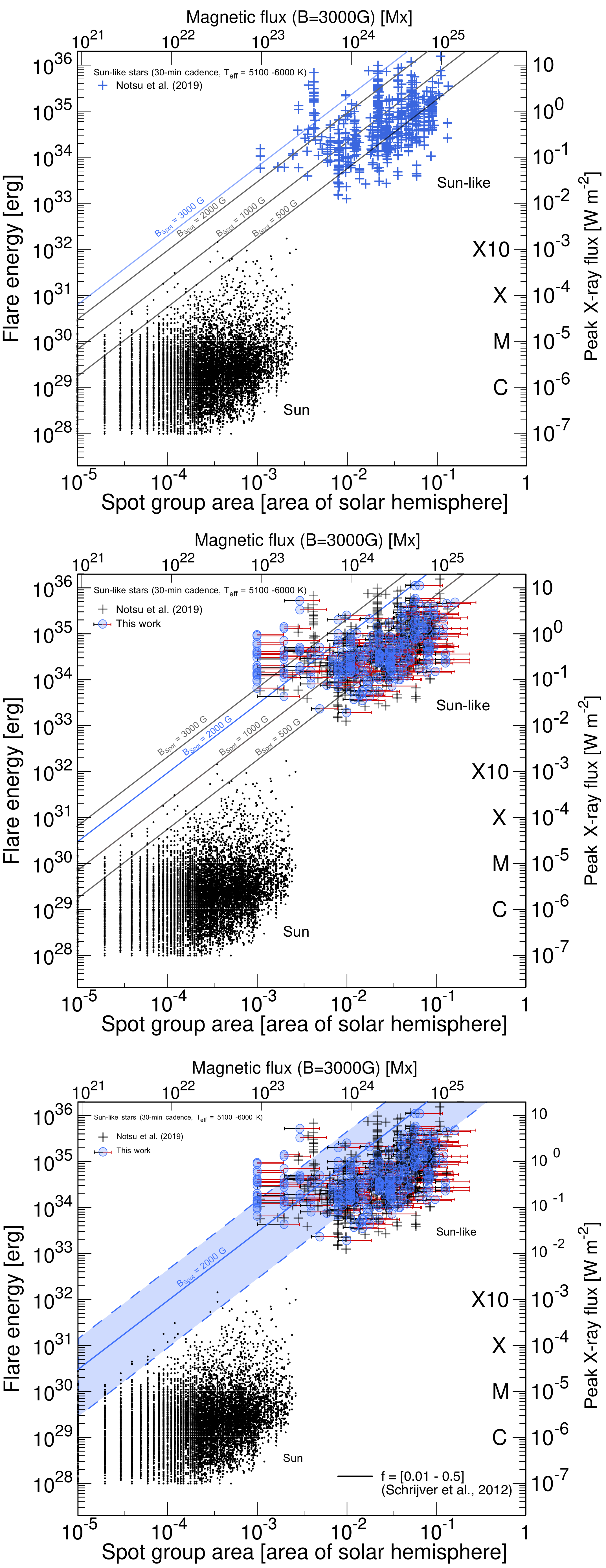}
  \caption{Upper panel: Scatter plot of the flare energy as a function of spot area for solar flares (black dots) and superflares of G-type stars \citep{Notsu-etal-2019}. Middle panel: Updated stellar sample by \citet{Notsu-etal-2019} based on the newly-derived $\frac{\Delta F}{F_{\mathrm{Star}}}$. 
  Lower panel: Influence of the utilized $f$-values. Upper and lower limits according to \citet{Schrijver-etal-2012} [f = 0.01 - 0.5].}
 \label{fig:3}
 \end{center}
 \end{figure}

\begin{figure}[!t]
\begin{center}
    \includegraphics[width=0.48\textwidth]{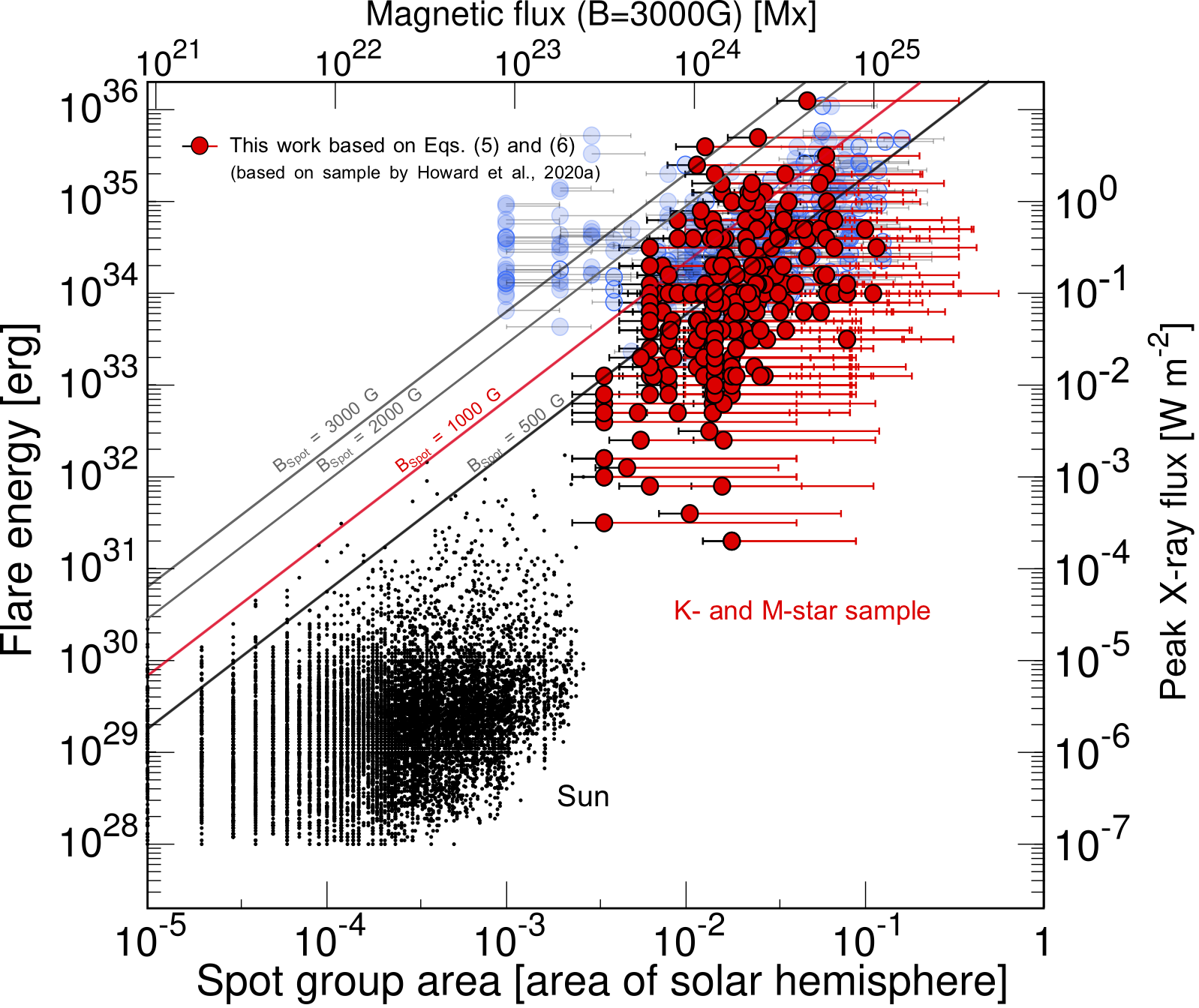}\\
    \includegraphics[width=0.48\textwidth]{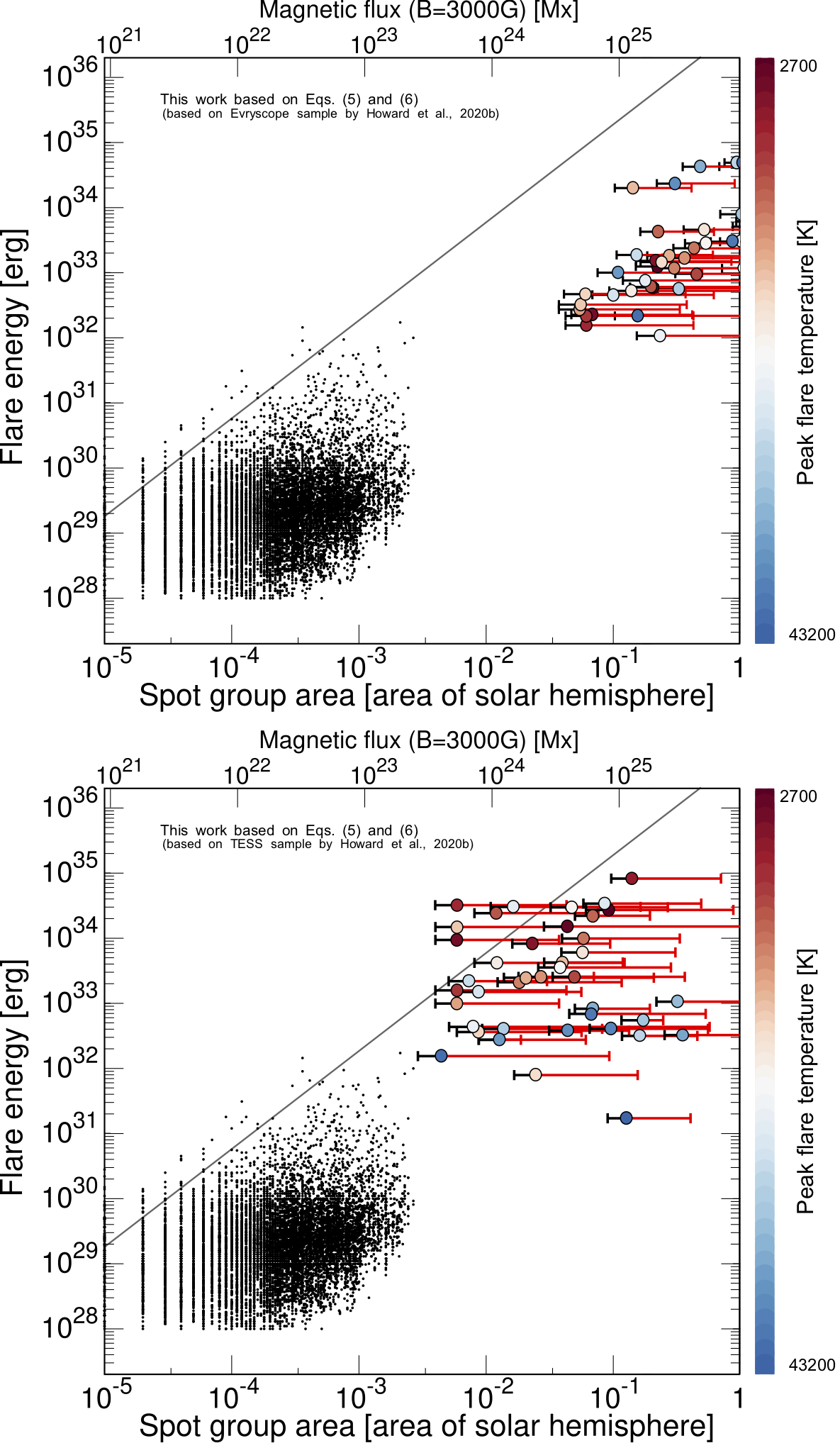}
  \caption{Upper panel: Updated K- and M-star sample of the narrow-band observations of the \textit{Evryscope} instrument \citep{Howard-etal-2019b}. Middle panel: 42 superflares of K5-M5 dwarfs most recently observed with the Evryscope instrument \citep{Howard-etal-2020}. Color-coding corresponds to the measured peak flare intensities. Lower panel: Same superflares observed by TESS.}
 \label{fig:4}
 \end{center}
 \end{figure}

Stellar flares are sudden releases of magnetic energy that is stored in magnetic active regions. According to \citet{Sturrock-etal-1984}, \citet{Shibata-etal-2013} and \citet{Notsu-etal-2019}, the upper limit of the total released flare energy $E_{\text {flare }}$ can be described as constant fraction of the total non-potential (stressed) magnetic field energy of the active region \citep[see][]{Emslie-etal-2012}, and thus as a rather simple scaling-law in the form of
\begin{equation}
\begin{aligned}
E_{\text {flare }} & \approx f E_{\text {mag }} \\
&\approx f \frac{B^{2}}{8 \pi} A_{\text {spot }}^{3 / 2}\label{eq:eflare1}
\end{aligned}
\end{equation}
and thus
\begin{equation}
\begin{aligned}
E_{\text {flare }} & \approx a\left(\frac{f}{0.1}\right)\left(\frac{B}{10^{3}~\rm{G}}\right)^{2}\left(\frac{A_{\text {Spot }} /(2\pi R_{\odot}^2)}{0.001}\right)^{3 / 2},\label{eq:eflare}
\end{aligned}
\end{equation}
where $a$ = 7 $\times$ 10$^{32}$ erg, $f$ gives the fraction of magnetic energy released as flare energy, and $B$ represents the magnetic field strength of the solar/stellar active region (AR) given in Gauss. It becomes apparent that the derived spot temperature is a critical parameter for determining the flare energy plotted in the panels of Figs.~\ref{fig:3} and \ref{fig:4}. Thereby, $f$ often is assumed to be in the order of 10\%, with upper limits reaching 50\% \citep[see][]{Schrijver-etal-2012} or even higher \citep[see][]{Aschwanden-etal-2014}. 

In addition, many years of research on our host star have shown that ARs producing intense flares tend to have a larger spot area as well as morphological and magnetic complexity. Large amounts of magnetic energy are accumulated in ARs; hence, it is reasonable to be larger in the spot area. At the same time, the morphological and magnetic complexity of an AR is manifested during its evolution and serves as a trigger of eventual flare eruption \citep[for a comprehensive review see][]{2019LRSP...16....3T}. Further, a multi-wavelength multi-observatory Sun-as-a-star study on the spectral irradiance observations of transiting ARs has been performed by \citet{Toriumi-etal-2020}. Since the complexity and wealth of observational evidence concerning ARs can not be captured, Eqs.~(\ref{eq:eflare1}) and (\ref{eq:eflare}) can only be considered as simple but yet valuable approximations that provide a baseline for the construction of empirical relations of stellar ARs and flares. Several studies \citep[e.g.][]{Basri-Nguyen-2018} emphasize that it is not clear whether the brightness modulation amplitude is related to the size of superflare-generating starspots. However, most recently, \citet{Namekata-etal-2020} showed that the area estimated from the rotational modulation is consistent with the maximum in-transit spots, indicating that the Kepler light-curve amplitude is a good indicator of the maximum visible spot size, and thus is a good proxy for the energy released in a flare.

Thus, keeping in mind that the contours reflect the maximum energy release limit, Eq.(\ref{eq:eflare}) can be used to study the influence of the magnetic field strength of starspots on the energy released in a flare. The first two panels of Fig.~\ref{fig:3} show the corresponding contours for B-values of 3000~G, 2000~G, 1000~G, and 500~G, respectively. As can be seen in the upper panel of Fig.~\ref{fig:3}, the majority of the solar observations \citep[black dots, from][]{Sammis-etal-2000} can be found below the $B_{\rm{Spot}}$ = 500~G contour line when $f=$ 0.1 is assumed. Note, however, that according to \citet{Aulanier-etal-2013}, these contour lines only account for the outer spot area. Correspondingly, the represented B-values have to be multiplied by a factor of five to reflect the magnetic field strength of the spot itself. With that, the solar observations point to an upper limit of $B_{\rm{Spot}}$ = 2500 G, which is in good agreement with observations that show $B_{\rm{Spot}}$ to vary between 1500 G and 3600 G \citep[][]{Livingston-etal-2015}. However, a small number of observations indicate the existence of higher $B_{\rm{Spot}}$-values, which might be a direct indicator that the solar dynamo is capable of causing a higher maximum energy release limit. The latter further puts emphasis on the existence of strong solar events as detected in the cosmogenic radionuclide records of $^{10}$Be, $^{14}$C, and $^{36}$Cl around AD774/775, AD 993/994, and BC660 \citep[][respectively]{Miyake-etal-2012, Mekhaldi-etal-2015, OHare-etal-2019}.
%
%
%
Additionally, the upper panel of Fig.~\ref{fig:3} shows the stellar Sun-like sample by \citet{Notsu-etal-2019} (blue crosses). As discussed by \citet{Notsu-etal-2019}, most of the stellar superflares are located below the 3000 G line. According to \citet{Notsu-etal-2019}, stars located above this line are considered to have either a low inclination angle or starspots around the poles. However, by taking into account our updated stellar spot-sizes and the limiting errors, as shown in the middle panel, this picture slightly changes. In particular, by taking into account the upper spot group area limits (red error bars) almost all stars of the G-type star sample are found to be well below the 3000 G line, with the majority of stars even located below the 2000 G line, which strongly suggests that the upper limit of the energy released by the observed flares does not exceed the threshold of the magnetic energy stored around the starspots. However, since the utilized $f$-values have a strong impact on the estimated upper flare energy limits, with values discussed in literature ranging from 0.01 to 0.5 and higher, Eqs.~(\ref{eq:eflare1}) and  (\ref{eq:eflare}) represent only very rough upper estimates (see the lower panel of Fig.~\ref{fig:3}).

In the case of the K- and M-star regime observed with the narrow-band \textit{Evryscope} instrument \citep{Howard-etal-2019b}, however, the differences between the previously published and the newly-derived upper limits of the flare energy release are more severe (see upper panel of Fig.~\ref{fig:4}). According to \citet{Howard-etal-2019b} the entire sample lies to the right of the 2000~G line, while the majority can be found to the right of the 500 G contour. Based on Eq.~(\ref{eq:6}), however, this picture changes. As can be seen cool stars with $B$-values above the 3000~G line can be found, while the majority of stars is located to the right of the 1000~G line, which is in good agreement with both the measured magnetic field strengths of cool stars \citep[above 4000 G, e.g.][]{Shulyak-etal-2017, Shulyak-etal-2019} as well as the updated G-type sample by \citet{Notsu-etal-2019}.

For the Sun-like stellar sample discussed previously, \citet{Notsu-etal-2019} follow \citet{Shibayama-etal-2013} to estimate the total energy of each flare from the stellar luminosity, amplitude, and duration of flares by assuming that the spectrum of white-light flares can be described by a blackbody radiation with an effective flare temperature of $T_{\rm{flare}}$ = 10,000~K \citep[][]{Mochnacki-Zirin-1980, Hawley-Fisher-1992}. However, \citet{Howard-etal-2020} most recently published multi-band observations of 42 superflares from 27 K5 - M5 stars. 
For the first time, a systematic exploration of the superflare temperature evolution is provided, showing constrained blackbody temperatures of flares ranging above 14,000 K up to 42,000 K. While the middle panel of Fig.~\ref{fig:4} displays the observations of the narrow-band Evryscope instrument and the corresponding estimations based on Eq.~(\ref{eq:5}) and (\ref{eq:6}), the lower panel shows the results for the simultaneously observed superflares by TESS \citep[broad-band observations, data taken from][]{Howard-etal-2020}. However, these values should be treated with caution since \citet{Howard-etal-2020} did not consider that the optical emission from K- and M-star flares usually deviate from blackbody spectra \citep[e.g., Fig. 16 of][]{Kowalski-etal-2019} because of the significant contribution from chromospheric emission lines in the visible band (e.g., hydrogen Balmer lines and He I).

Comparing both samples shows that the energies released in the superflares are comparable between the bandpasses. However, because Evryscope uses a narrow bandpass in "blue" wavelength (4000-5500\AA), where the contrast between spot and photosphere is larger than in broad-band observations, the observed amplitudes are in the order of one magnitude higher, and so are the observed $\Delta T/T$-values and the estimated spot group areas.

\section{From Stellar Flares to Coronal Mass Ejections}
The total flare energy can further be scaled with the energy released in an associated CME. It is known, that fast and energetic CMEs induce shocks in the corona that serve as sites of gradual SEP events with the flux of accelerated protons to energies of a few GeV \citep[][]{Gopalswamy-etal-2005, Fu-etal-2019}. 

According to \citet{Takahashi-etal-2016} the CME mass ($M_{\rm{CME}}$), the CME speed ($v_{\rm{CME}}$), and the corresponding released energetic peak proton flux ($F_p$) are related to $E_{\rm{flare}}$ based on simple power-law relations. However, three assumptions have to be made: (1) $M_{\rm{CME}}$ is defined as the sum of all mass within the gravitationally stratified coronal active region, (2) the kinetic energy of the CME is proportional to $E_{\mathrm{flare}}$, and thus
\begin{equation}
E_{\mathrm{CME}} = \frac{1}{2} M_{\mathrm{CME}} V_{\mathrm{CME}}^{2} \propto E_{\mathrm{flare}} = f \frac{B^{2}}{8 \pi} A_{\text {spot }}^{3 / 2}, 
\end{equation}
Therefore, the following dependencies can be derived: $v_{\mathrm{CME}}\propto E_{\rm{flare}}^{1/6} \propto F_{\rm{SXR}}^{1/6}\propto F_p^{1/5}$, where  $F_{\rm{SXR}}$ is the soft X-ray flux and  $F_{\rm{p}}$ the stellar proton flux released in a flare.  Thus, $F_p$ can be assumed to be proportional to $E_{\rm{flare}}^{5/6}$, and therefore
\begin{equation}
F_p \propto \left(4.427\times 10^{34} \cdot f \cdot B \cdot \left(\frac{A_{\rm{Spot}}}{2\pi R_{\odot}^2}\right)^{3/2}\right)^{5/6}.
\end{equation}
Consequently, based on the latter, the proton flux of particularly M-stars utilizing the previously reported upper magnetic field strength values would be underestimated by about 40\%.
%
\section{Summary and Conclusions}
The motivation of this study was to investigate the emperical relation between stellar effective temperatures and the temperatures of starspots and the relations for eruptive events based on it.

To this end we utilized the stellar sample by \citet{Berdyugina-2005}, and proposed a new empirical relation (Eqs.~(\ref{eq:3}) and (\ref{eq:4})), for the first time providing a corresponding error-band.

Further, the original stellar sample has been extended. By utilizing the measurements of K- and G-type star sample by \citet{Biazzo-etal-2006} and the modeled values of the K-, F-, and G-type star sample by \citet{Valio-2016} a new semi-analytic relation is derived that can be applied for all cool star types (Eqs.~(\ref{eq:5}) and (\ref{eq:6})). As a result, the newly-derived $\Delta T(T_{\mathrm{Star}})$-values vary between 500 K in the case of cool M-stars (M4) up to about 1500 K in the case of G-type stars (G0). Consequently, the higher-contrast of spots with respect to the photosphere of hotter stars not necessarily is as pronounced as previously thought.

The difference between stellar effective temperatures and starspot temperatures further is often used to derive the size of starspots on the stellar surface \citep[see, e.g.,][]{Notsu-etal-2013, Maehara-etal-2017, Notsu-etal-2019, Howard-etal-2019b}. It showed that the average difference of the derived $A_{\rm{Spot}}/A_{\rm{Star}}$-values based on the former relations by \citet{Berdyugina-2005} and the newly-derived relation discussed in this study is in the order of 20\% in the case of G-type stars while being up to 40\% in the case of M-type stars, which implies that previous studies most likely underestimated the starspot sizes of cool stars.

In addition, the magnetic energy stored around starspots is often released in the form of stellar flares. According to \citet{Sturrock-etal-1984}, \citet{Shibata-etal-2013} and \citet{Notsu-etal-2019}, the upper limit of the total energy released in a flare $E_{\text {flare}}$ can be described as constant fraction of the magnetic field energy of the active region \citep[see][]{Emslie-etal-2012}, and thus as a rather simple scaling-law. For the G-type star sample reported by \citet{Notsu-etal-2019}, we found that the stellar magnetic fields of at least 2000 G are most consistent with the measurements. The latter reduces the previously published upper limit by about 1000 G. In addition, \citet{Howard-etal-2019b} found that stellar magnetic fields of at least 500 G are most consistent with the narrow-band observations of the \textit{Everyscope} instrument. Applying the newly-derived correlations, in this study we found that the upper limit of the M-star flare sample most likely is around 1000 G.

We further note, that flares with much lower energy releases are to be expected, however, are not present in the observations due to detection limitations. Solar observations by the \textit{Sphinx} instrument (about 100 times more sensitive than the GOES instrument), for example, lead to the detection of flares much smaller than the previously known flare classes \citep[S and Q-class flares, see][]{Gryciuk-etal-2017}. In addition, recent \textit{TESS} observations of the M-star GJ 887 showed multiple M- to X-class flares \citep[][]{Loyd-etal-2020} often observed at the Sun.

Nevertheless, the relationships we proposed in this study require further observations (e.g., using instruments like the \textit{XMM-Newton} and \textit{NICER}) and numerical modeling of starspot temperatures for G-M dwarfs, which allow an extended empirical characterization of the stellar relationships using all available observational data at hand. The latter will be subject of future investigations.
%
\section*{Acknowledgements}
V.S.A. would like to acknowledge the TESS Cycle 1 support.  
D.A. acknowledges support by the New York University Abu Dhabi (NYUAD) Institute research grant G1502. 
The authors further gratefully acknowledge the International Space Science Institute and the supported International Team 464: \textit{The  Role  Of  Solar  And  Stellar  Energetic Particles On (Exo)Planetary Habitability (ETERNAL, \url{http://www.issibern.ch/teams/exoeternal/})}. 

\bibliographystyle{aasjournal}
\bibliography{references}
\end{document}